\documentclass[submission,copyright,creativecommons]{eptcs}
\providecommand{\event}{PROLE 2015} 
\usepackage{breakurl}             
\usepackage{amsthm}
\usepackage{amsmath}
\usepackage{proof}
\usepackage{amssymb}
\usepackage{graphicx}
\usepackage{ifthen}
\usepackage{array}
\usepackage{setspace}

\title{Restricted Predicates for Hypothetical Datalog}
\author{Fernando S\'aenz-P\'erez\thanks{Work partially supported by the Spanish MINECO project CAVI-ART (TIN2013-44742-C4-3-R), Madrid regional project N-GREENS Software-CM	(S2013/ICE-2731) and UCM grant GR3/14-910502.}
\institute{Facultad de Inform\'atica\\
Universidad Complutense de Madrid\\
Madrid, Spain}
\email{fernan@sip.ucm.es}
}
\def\titlerunning{Restricted Predicates for Hypothetical Datalog}
\def\authorrunning{F. S\'aenz-P\'erez}

\newcommand{\myfontcodesize}{\fontsize{11}{11}}
\newcommand{\mytt}[1]{{\myfontcodesize \texttt{#1}}}

\def\smallskip{\vskip\smallskipamount}
\def\medskip{\vskip\medskipamount}
\def\bigskip{\vskip\bigskipamount}

\newcounter{example}

\def\exclaim#1{\par\medskip\noindent\refstepcounter{example}\hbox{\em Example \arabic{example} 
        \ifthenelse{\equal{#1}{}}{}{(#1):}
    }
}
\def\endexclaim{\hfill$\square$
    \par\medskip}
\newenvironment{example}{\exclaim}{\endexclaim}

\newcounter{mydef}

\def\claim#1{\par\medskip\noindent\refstepcounter{mydef}\hbox{\em Definition \arabic{mydef} 
        \ifthenelse{\equal{#1}{}}{}{(#1)}
    }
}
\def\endclaim{
    \par\medskip}
\newenvironment{mydef}{\claim}{\endclaim}

\newcounter{mylemma}

\def\lclaim#1{\par\medskip\noindent\refstepcounter{mylemma}\hbox{\em Lemma \arabic{mylemma}.}
    \it\ 
}
\def\endlclaim{\hfill$\square$
    \par\medskip}

\begin{document}
\maketitle

\begin{abstract}
Hypothetical Datalog is based on an intuitionistic semantics rather than on a classical logic semantics, and embedded implications are allowed in rule bodies.
While the usual implication (i.e., the neck of a Horn clause) stands for inferring facts, an embedded implication plays the role of assuming its premise for deriving its consequence.
A former work introduced both a formal framework and a goal-oriented tabled implementation, allowing negation in rule bodies.
While in that work positive assumptions for both facts and rules can occur in the premise, negative assumptions are not allowed.
In this work, we cover this subject by introducing a new concept: a restricted predicate, which allows negative assumptions by pruning the usual semantics of a predicate.
This new setting has been implemented in the deductive system DES.
\end{abstract}

\section{Introduction}
\label{sect:intro}
Hypothetical queries are a common need in several scenarios, related mainly with business intelligence applications and the like.
They are also known as "what-if" queries and help managers to take decisions on scenarios which are somewhat changed with respect to a current state.
Such queries are used, for instance, for deciding which resources must be added, changed or removed to optimize some criterium (i.e., a cost function, a notion well related to optimization technologies).
Current applications include OLAP environments \cite{DBLP:conf/bife/ZhouCZ09}, business intelligence \cite{DBLP:journals/jdwm/GolfarelliR09a}, and e-commerce \cite{Zhang:2007:AHQ:1304611.1306565}.
Even, major vendors of relational databases include (quite limited) approaches to hypothetical queries, as for instance the model clause in Oracle SQL data warehousing \cite{oracle:data:warehousing:2011}.

Whilst such systems and applications inherit from and build upon relational databases and restrict the use of negation and recursion, earlier works on logic programming fully integrate hypothetical queries in the inference system.
These approaches \cite{DBLP:journals/jlp/McCarty88,Miller86atheory,DBLP:journals/jlp/Gabbay85} fit into intuitionistic logic programming, an extension of logic programming including both embedded implications and negation.
In particular, Hypothetical Datalog \cite{Bonner89hypotheticaldatalog,bonner90adding} has been a proposal thoroughly studied from semantic and complexity point-of-views.

A recent work on tabled Hypothetical Datalog \cite{sae13c-ictai13} extended \cite{bonner90adding} by adding a number of extensions: First, allowing to include rules in embedded implication premises, with the intention to allow the user to assume not only facts but also rules. Second, support for duplicates allowing multiple copies of the same tuple, whose source can be either extensional (a bag of facts) or intensional (rules delivering such multiple copies), which in addition can be summarized with aggregates (as counting them). And, finally, support for strong integrity constraints which enable to reject rules and facts which do not meet the integrity criterion (in the same line as relational databases do).
However, while \cite{sae13c-ictai13} allows to locally add tuples (in the context of an embedded implication) to the database, it lacks the ability to locally delete tuples in the same context.
In \cite{Bonner90hypotheticaldatalog} support for such deletions are provided, though only for facts.


In this paper, we extend \cite{sae13c-ictai13} by allowing deletions of tuples, not only for facts as data providers as in \cite{Bonner90hypotheticaldatalog}, but also for rules.
Then, rules can be used to intensionally specify those facts that must be deleted in a given context.
To this end, we introduce the novel concept of restricted predicates, which include usual facts (extensional specification) and rules (intensional specification) --which we refer to as {\em positive} from now on-- along with restricted versions of them --which we refer to as {\em negative} from now on--.
So, whereas additions are captured with usual predicates, deletions are captured with restricted predicates.
All the features in \cite{sae13c-ictai13} that extended \cite{bonner90adding} are preserved in this new deletion setting.
In both cases, two kind of implications are identified: The usual implication ($\leftarrow$) which is found as the neck of a logic clause, and the (hypothetical) intuitionistic implication ($\Leftarrow$) which can be found in the body of a logic clause.
Note that intuitionistic implication is not transitive as the classical logic implication \cite{bonner90adding}.

We have implemented this proposal in the deductive system DES (\url{des.sourceforge.net}), completing the implementation described in \cite{sae13c-ictai13} with support for negation in bodies and restricted predicates in premises.
Though there have been some works regarding implementations \cite{DBLP:conf/lpar/VieilleBKLM92}, as far as we know there has not been an implementation of hypothetical Datalog with intensional deletions.
%

With respect to related work, Date \cite{Date2009} explains the idea behind such ``what-if" statements, an approach that was firstly proposed in \cite{Stonebraker:1980:EEK:582250.582261} for relational databases.
A recent work \cite{EasyChair:355} also develops this idea by generating database scripts that implement a fixpoint computation for building SQL materialized views as tables.
In \cite{DBLP:conf/dagstuhl/ChristiansenA98}, an approach to hypothetical database query evaluation based on counterfactual reasoning is proposed.
Though it includes both positive and negative assumptions, it only includes these in queries, but not in rules.
In the logic programming field, Miller and Nadathur worked at developing and justifying the intuitionistic theory of hereditary Harrop (HH) formulas (see, e.g., \cite{uniform}), which lead to the implementation of $\lambda$Prolog.
A more recent work \cite{ANSS14jlap} proposes $HH_\neg(C)$ as a constraint database framework including negation, but with no negative assumptions.

Organization of this paper proceeds as follows: Section \ref{sect:example} introduces an example to show that negative assumptions can be handy to solve some queries.
Section \ref{sect:restricted} introduces some examples illustrating assumptions and the notion of restricted predicate as the device to capture negative assumptions in premises.
Next, some formal background is presented in Section \ref{sect:formal}.
The setting to implement this background is described in Section \ref{sect:implementation} as part of the deductive system DES.
Finally, Section \ref{sect:conclusions} concludes and lists some future work.

\section{Introductory Example}
\label{sect:example}
With respect to logic programming, in the context of deductive databases, the term {\em relation} is used interchangeably with {\em predicate}, {\em rule} with {\em clause} and the term {\em query} with {\em goal}.
Also, we identify two components of a deductive database: The extensional database (EDB) which is composed of predicates defined only by facts, and the intensional database (IDB) which is composed of predicates defined at least by one rule.
From now on, all examples written in true type can be actually run in DES.
Before introducing the example and others in next sections, we recall the concrete syntax of hypothetical queries, extending the premise to include {\em restricting rules}.
These restricting rules will be useful for answering questions with embedded implications that are neither possible in \cite{Bonner90hypotheticaldatalog} nor in \cite{sae13c-ictai13}.

\vspace*{-2.5mm}
\subsection{Concrete Syntax}
The syntax of a hypothetical query in the system DES is as follows:
{\myfontcodesize
    \begin{verbatim}
    rule1 /\ ... /\ ruleN => goal
    \end{verbatim}
}
\vspace*{-5mm}
\noindent where each rule \mytt{rule{\em i}} can be a regular (usual) rule or a restricting rule.
A restricting rule has a head of the form \mytt{-Atom}, where \mytt{Atom} is an atom.
For facts, the body is empty (no neck symbol either) and the atom is ground to ensure safety \cite{Ullman88} (rules and queries must be safe as well).
Such a hypothetical query represents that, assuming that the current database is augmented with the regular rules in $\mathcal{R}=$\{\mytt{rule{\em i}} $|$ $\mytt{1}\leq \mytt{{\em i}} \leq \mytt{N}$\}, 
and that the meaning of the restricting rules in $\mathcal{R}$ are removed from their corresponding predicates,
then \mytt{goal} is computed with respect to such modified current database.
Note that the implication symbol \mytt{=>} (intuitionistic implication to the right) is used for the so-called embedded implication in {\em lieu} of the classical implication \mytt{:-} (implication to the left) typically used in logic programming systems.

Such query is also understood as a literal in the context of a rule, so that any rule can contain hypothetical goals (in particular, any \mytt{rule{\em i}}). 
Variables in each \mytt{rule{\em i}} are encapsulated w.r.t. the rule (i.e., they are neither shared with other rules nor with the goal, even when they might have the same name).
Moreover, a hypothetical literal does neither share variables with other literals nor with the head of the rule in which it occurs.

As it is usual in logic programming systems, variables start with upper case or underscore and other program identifiers either start with lower case or are delimited by single quotes.

\subsection{A University Example}
Borrowing an example from \cite{Bonner90hypotheticaldatalog}, we consider an extended and adapted rule-based system for describing a university policy.
EDB is composed of: \mytt{student(S)} (meaning that \mytt{S} is a student), \mytt{course(C)} (\mytt{C} is a course), and \mytt{take(S,C)} (student \mytt{S} takes course \mytt{C}).
And IDB is: \mytt{grad(S)} (student \mytt{S} is eligible for graduation).
EDB contains facts as:

{\myfontcodesize
\begin{verbatim}
student(adam). student(scott).  course(eng).  take(adam,eng). take(scott,his).
student(bob).  student(tony).   course(his).  take(pete,his). take(scott,lp).
student(pete).                  course(lp).   take(pete,eng). take(tony,his).                             
\end{verbatim}
}

IDB can contain rules as: \mytt{grad(S) :- take(S,his), take(S,eng).}
%

A regular query for students that would be eligible to graduate is:

{\myfontcodesize
\begin{verbatim}
DES> grad(S)
{ grad(pete) }
\end{verbatim}
}

\noindent where the answer is the bag of goal instances delimited between curly brackets.

\begin{example}{}
A first hypothetical query for this database asks "If Tony took \mytt{eng}, would he be eligible to graduate?": 
{\myfontcodesize
\begin{verbatim}
DES> take(tony,eng) => grad(tony)
Info: Processing:
  answer :- take(tony,eng)=>grad(tony).
{ answer }
\end{verbatim}
}
\noindent Here, the query has been automatically rewritten as a temporary view with name \mytt{answer}, i.e., a view which is added to the database and eventually removed.
This allows non atomic goals to be solved, as it is the case for an implication.
The outcome of the query is the result of the goal \mytt{answer}, which can be proved because assuming that premise allows to deduce the consequent.
\end{example}

\begin{example}{}
Also, more than one assumption can be simultaneously stated, as in: "If Tony took \mytt{eng}, and Adam took \mytt{his}, what are the students that are eligible to graduate?": 
{\myfontcodesize
\begin{verbatim}
DES> take(tony,eng) /\ take(adam,his) => grad(S)
Info: Processing:
  answer(S) :- take(tony,eng)/\take(adam,his)=>grad(S).
\end{verbatim}
\vspace*{-1.8mm}
\noindent
\verb+{ answer(adam), answer(pete), answer(tony) }+
}
\end{example}

\pagebreak

\begin{example}{}
Another query is "Which are the students which would be eligible to graduate if \mytt{his} and \mytt{lp} were enough to get it?": 
{\myfontcodesize
\begin{verbatim}
DES> (grad(S) :- take(S,his), take(S,lp)) => grad(S)
Info: Processing:
  answer(S) :- (grad(S):-take(S,his),take(S,lp))=>grad(S).
{ answer(pete), answer(scott) }
\end{verbatim}
}
Note that, although \mytt{S} occurs in both the premise and the conclusion of \mytt{=>}, they are not actually shared, and they simply act as different variables.
\end{example}

\begin{example}{}
\label{ex:assum-noassum}
Let us consider the following question: ``Which are the new students that are eligible to graduate if we consider that \mytt{his} and \mytt{lp} were enough to graduate?" This query needs to compare the students under the assumption with the students with no assumption at all. 
A possible formulation is:
{\myfontcodesize
\begin{verbatim}
DES> ((grad(S) :- take(S,his), take(S,lp)) => grad(S)), not grad(S)
Info: Processing:
  answer(S) :- (grad(S):-take(S,his),take(S,lp)) => grad(S)),not grad(S).
{ answer(scott) }
\end{verbatim}
}
Note that the assumption affects only to the first goal \mytt{grad(S)}.
This assumption does not affect to the second, negated goal \mytt{grad(S)}.
Negation allows to compute the set difference of students.
\end{example}
    
\begin{example}{}
Next rules represent information about course prerequisites:
{\myfontcodesize
\begin{verbatim}
pre(eng,lp).  pre(hist,eng).
pre(Pre,Post) :- pre(Pre,X), pre(X,Post).
\end{verbatim}
}
Whether adding a new prerequisite implies a cycle can be asked with: 
{\myfontcodesize
\begin{verbatim}
DES> pre(lp,hist)=>pre(X,X)
Info: Processing: answer(X) :- pre(lp,hist)=>pre(X,X).
{ answer(eng), answer(hist), answer(lp) }
\end{verbatim}
}
The answer includes those nodes in the graph that are in a cycle.

Another option is to avoid cycles by using the following strong constraint (which are defined in \cite{sae13c-ictai13}):

{\myfontcodesize
	\begin{verbatim}
	DES> :-pre(X,X)
	\end{verbatim}
}

\noindent which means that it should not be the case of finding a subject that depends on itself. 
Then, to list prerequisites assuming {\tt pre(lp,hist)}:

{\myfontcodesize
	\begin{verbatim}
	DES> pre(lp,hist)=>pre(X,Y)
	Info: Processing:
	answer(X,Y) :- pre(lp,hist)=>pre(X,Y).
	Error: Integrity constraint violation.
	ic(X) :- pre(X,X).
	Offending values in database: [ic(lp),ic(eng),ic(hist)]
	Info: The following rule cannot be assumed:
	pre(lp,hist).
	{ answer(eng,lp), answer(hist,eng), answer(hist,lp) }
	\end{verbatim}
}

So, the system informs that there is an inconsistency when trying to assert such offending fact (\mytt{pre(lp,hist)}), which makes prerequisites to form a cycle (as shown in the offending value list \mytt{[ic(lp),ic(eng),ic(hist)]}).
The system informs about the rules that cannot be assumed but continues its processing.
This is also useful to know the result for the admissible assumptions.
Note that, in general, offending facts can be a subset of the meaning of an assumed rule in the context of the current database.
To illustrate this, let's consider a game that students like to play that consists of tossing a coin:

{\myfontcodesize
\begin{verbatim}
% Tails win:
:- win, heads.
win :- heads ; tails.
\end{verbatim}
}

The predicate {\tt win} states that one wins if either heads or tails are got, and the constraint states that you have to get tails to win.
Here, the semicolon ``;" denotes disjunction as in Prolog syntax.
Then, the following hypothetical goal states whether assuming heads or tails leads to win.

{\myfontcodesize
\begin{verbatim}
DES> heads /\ tails => win
Info: Processing:
answer :- heads/\tails=>win.
Error: Integrity constraint violation.
ic :- win, heads.
Info: The following rule cannot be assumed: heads.
{ answer }
\end{verbatim}
}

As it is informed, heads cannot be assumed in order to win.
\end{example}

\begin{example}{}
\label{ex:unsolved1}
Consider a query as: "If Pete had {\em not} taken \mytt{eng}, could he have graduated?", which is equivalent to say: "If \mytt{take(pete,eng)} were deleted from the database, could we infer \mytt{grad(pete)}?" 
This query cannot be solved with the former proposal in \cite{sae13c-ictai13} but is supported in \cite{Bonner90hypotheticaldatalog}.
\end{example}

\begin{example}{}
\label{ex:unsolved2}
Further, consider the query: "What would happen if the current prerequisites were the other way round?" 
This would imply to remove the intensional rule about prerequisites and add a modified one.
In turn, this is neither supported by \cite{Bonner90hypotheticaldatalog} nor by \cite{sae13c-ictai13}.
\end{example}

The next section introduces restricted predicates as a means to provide semantics to such deletions, which are referred to as {\em negative assumptions} in the context of an embedded implication.
Then, it will be possible to specify these last two examples with such implications.

\section{Restricted Predicates: Informal Semantics}
\label{sect:restricted}
Here, we introduce the novel concept of {\em restricted predicate}.
The intention is to prune the meaning of a usual predicate by specifying some {\em restricting rules}.
A restricting rule is a rule for which its head is a {\em restricting atom} (a regular atom preceded by a minus sign \mytt{-}). 
We use the term {\em regular rule} to refer to a rule which is not a restricting rule (i.e., usual Horn logic rules).

The meaning of a restricted predicate is then the tuples deduced from its regular rules minus the tuples deduced from its restricting rules. 
Note that a restricting rule does not represent true negation, but a means to discard positive tuples from the meaning of a predicate. 
So, both \mytt{p} and \mytt{-p} can occur in a program with no contradiction at all in a single model.
By contrast, this situation in classical negation results in contradiction \cite{Gelfond91classicalnegation}.

In our setting, computing a restricted predicate \mytt{p} can be roughly seen as follows: First, compute its meaning $P^+$ from its regular rules. 
Then, compute the meaning $P^-$ of its restricting rules and build the meaning for \mytt{p} as the difference $P^+ - P^-$. 
As it will formalized in Section \ref{sect:pdg}, adding a restricting rule for a predicate involves to add a negative dependency \mytt{q}$\stackrel{\neg}{\leftarrow}$\mytt{p} from any other predicate \mytt{q} depending on \mytt{p}.
This implies that such other predicate \mytt{q} will be located in a higher stratum than \mytt{p}.
Therefore, from an operational point-of-view, the meaning of \mytt{p} must be computed before that of \mytt{q}.
This ensures monotonicity along fixpoint computation as it will not be the case of considering a given tuple in a meaning that can be discarded afterwards in another iteration cycle.
This is a similar requirement as done for stratified negation \cite{Ullman88} and will be formalized in Section \ref{sect:formal}.

Bearing this in mind, we can think of the next example.
Let us consider the following number generator:

{\myfontcodesize
\begin{verbatim}
DES> /assert p(X) :- X=1 ; p(Y), Y<10, X=Y+1.
DES> p(X)
{ p(1), p(2), ..., p(10) }
Info: 10 tuples computed.          
\end{verbatim}
}

In the first line, a disjunctive rule is added to the current database.
Then, the whole meaning of the predicate \mytt{p}/1 can be retrieved with the query \mytt{p(X)}.

Even numbers can be obtained by adding the following restricting rule to the current database:

{\myfontcodesize
\begin{verbatim}
DES> /assert -p(X) :- p(X), X mod 2 = 1.
DES> p(X)
{ p(2), p(4), p(6), p(8), p(10) }
Info: 5 tuples computed.          
\end{verbatim}
}

Now, the meaning of \mytt{p} as specified in the first assertion is changed by removing the tuples defined by the rule in the second assertion.
This way, atoms \mytt{p({\em i})}, with \mytt{{\em i}} odd, belong to the negative information of the program. 
That is, in particular it is possible to prove:

{\myfontcodesize
\begin{verbatim}
DES> not p(1)    
{ answer }
Info: 1 tuple computed.               
\end{verbatim}
}

Note that the definition of even numbers could be easily done with \mytt{p(X) :- X=2 ; p(Y), Y<10, X=Y+2.}
But this is not the point, what we are looking for is to {\em change} the meaning of a given predicate as shown later with the embedded implication.
This way, along a given query solving, the meaning of a given predicate can be changed with such an implication, while its meaning out of the consequence remains the same.
Example \ref{ex:assum-noassum} is an instance of this.

It is possible to inspect the meaning of the restricted part of a predicate ($P^-$ as introduced before). 
In general, a restricted atom can occur anywhere an atom is allowed, and, in particular, in a top-level query, as follows:

{\myfontcodesize
\begin{verbatim}
DES> -p(X)
{ -p(1), -p(3), -p(5), -p(7), -p(9) }
Info: 5 tuples computed.          
\end{verbatim}
}

And, conversely to the negation of the positive part of the program, we can ask if the negation of a restricted atom can be proven:

{\myfontcodesize
\begin{verbatim}
DES> not -p(1)    
{ }
Info: 0 tuples computed.               
\end{verbatim}
}

Summarizing, all the facts deduced from the restricted part of the program (either extensionally or intensionally) belong to the negative information of the program.

Restricting rules can also be recursive. 
The following example looks also for even numbers by removing odd numbers from \mytt{p}:

{\myfontcodesize
\begin{verbatim}
DES> /assert -p(X) :- X=1 ; -p(Y), X=Y+2, X<10.
DES> p(X)
{ p(2), p(4), p(6), p(8), p(10) }
Info: 5 tuples computed.          
\end{verbatim}
}

%
%

Coming back to the university example, the unsolved question in Example \ref{ex:unsolved1}
can now be posed as: 

{\myfontcodesize
\begin{verbatim}
-take(tony,eng) => grad(tony)
\end{verbatim}
}


Finally, the unsolved question in Example \ref{ex:unsolved2} can be posed as:

{\myfontcodesize
\begin{verbatim}
(-pre(Pre,Post) :- pre(Pre,X), pre(X,Post)) /\ 
( pre(Pre,Post) :- pre(Post,X), pre(X,Pre)) => pre(Pre,Post)
\end{verbatim}
}

%

\section{Formal Framework}
\label{sect:formal}
This section introduces some formal background to describe the approach to hypothetical Datalog we are considering, as an extension of function-free Horn logic following \cite{Bonner90hypotheticaldatalog,bonner90adding}.
Here, we recall and adapt the formal framework already presented in \cite{sae13c-ictai13},  presenting the syntax of the language, safety conditions, the notion of stratifiable program, and an operational semantics excerpt that extends \cite{sae13c-ictai13} with negative assumptions as restricting rules.
The main difference of the contents presented here w.r.t. \cite{sae13c-ictai13} is the inclusion of restricted predicates.

\subsection{Syntax}
The syntax of the logic is first order and includes a universe of constant symbols, a set of variables and a set of predicate symbols ($\mathcal{P}$).
For concrete symbols, we write variables starting with upper-case and the rest of symbols starting with lower-case.
Removing function symbols from the logic is a condition for finiteness of answers, a natural requirement of database users.
A rule has the form $A \leftarrow \phi$, where $A$ is either a regular atom or a restricting atom 
and $\phi$ is a conjunction of goals.
In addition, since we consider a hypothetical system, a goal can also take the form $G \leftarrow R$, a construction known as an embedded implication, where the premise $R$ represents an assumption and takes the form of a rule.
Moreover, we extend \cite{bonner90adding} by, first, allowing the premise to be a conjunction of rules $\bigwedge R_i$ as an assumption, and, second, allowing each $R_i$ to be a either a regular or a restricting rule.
From now on, we use the term rule to refer to both regular and restricting rules unless needed otherwise.

For solving the conclusion $G$, regular (restricting resp.) rules in $\bigwedge R_i$ will be used to augment (prune resp.) the meaning of their corresponding predicates with respect to the current database.
As an embedded implication behaves different from a regular implication \cite{bonner90adding}, it receives a different syntax symbol: $\Rightarrow$.
The following definition captures the syntax of the language, where $vars(T)$ is the set of variables occurring in $T$:

\begin{mydef}{Syntax of Rules}
\quad \\
$R := A \leftarrow G_1 \land \ldots \land G_n$ \\
$G := A \mid \neg G \mid R_1 \land \ldots \land R_m \Rightarrow G$\\
\noindent where $R$ and $R_i$ stand for rules (both regular and restricting), $G$ and $G_i$ for goals, $A$ for an atom (either regular or restricting), $n \geq  0$ (for $n=0$, $R$ is called a {\em fact}), $m>0$, and $vars(R_i)$ do not occur but in $R_i$.
\end{mydef}


Strong constraints are also supported in this new setting as rules with no head \cite{sae13c-ictai13}, and in the following we assume databases (as a set of rules and constraints) that are safe (with respect to query answers) and consistent (with respect to constraints) \cite{sae13c-ictai13}.

\subsection{Predicate Dependency Graph and Stratification}
\label{sect:pdg}

Introducing negation in literals of body clauses adds another issue: 
The possibility to have more than one minimal model \cite{Ullman88}.
Stratification is a syntactic condition on programs which ensures that only one minimal model can be assigned to a program.
Predicates in the program are classified into strata so that negation does not occur through recursion.
For building a stratification (i.e., a mapping between predicate symbols and natural numbers), a device called predicate dependency graph (PDG) is usually convenient.
A PDG depicts the positive and negative dependencies between predicates.

\begin{mydef}{Dependencies}
A predicate $P$ {\em positively} ({\em negatively}, resp.) depends on $Q$ if $P$ is the predicate symbol of $A$ in a rule (both a program rule and a rule in a premise) $A \leftarrow G_1 \land \ldots \land G_n$ and $Q$ occurs  either in some positive (either negative or restricting, resp.) atom $G_i$ or in $G$ in an embedded implication $G_j \equiv R_1 \land \ldots \land R_n \Rightarrow G$.
\end{mydef}

Note that the implication $\leftarrow$ is the source for dependencies, whereas the embedded implication $\Rightarrow$ is not.
However, all the non-atomic rules in the premise of $\Rightarrow$ are involved in adding dependencies.
This fact is propagated to the construction of the predicate dependency graph and the stratification for a program \cite{sae13c-ictai13}.
The PDG is the set of pairs $<N,A>$, where $N$ is the set of predicate symbols in $\Delta$ and $A$ is the set of arcs $P \leftarrow Q$ such that $P$ positively depends on $Q$, and $R \stackrel{\neg}{\leftarrow} S$ such that $R$ negatively depends on $S$.
The stratification is a mapping from predicates to integers such that if there is a dependency $R \leftarrow S$, then the integer assigned to $R$ must be less or equal than the one assigned to $S$. 
If the dependency is negative: $R \stackrel{\neg}{\leftarrow} S$, then the relation is strictly less.

%

\subsection{Stratified Inference}

Following \cite{bonner90adding} we define a logical inference system for stratified intuitionistic logic programming, with the following main differences: 
Allowing duplicates, integrity constraints, premises with multiple rules, and enforcing encapsulation of variables in premises.
Stratified inference requires an inference system for each stratum.
Inference starts from the lower stratum and its derivations are inputs to the inference for the next stratum above.
For a given stratum $i$, these derivations $\mathcal{A}$ are inference expressions which are constructed by the axioms derived in the stratum below and the rules defining the predicates belonging to stratum $i$.
Input $\mathcal{A}$ is the empty set for the first stratum.
In the following, we consider programs $\Delta$ which are both safe and stratifiable.
Otherwise, inference cannot be applied.

Duplicates would require working with bags (multisets) in order to denote the multiple occurrences of the same atom.
Instead, we resort to uniquely identifying each rule in a program and work with expressions tagged with such identifiers.

\begin{mydef}{Inference Expression}
\label{def:inference-expression}
An inference expression for a program $\Delta$ is denoted by $\Delta \vdash \psi$, where $\psi$ can be either an identified ground atom (either regular or restricting) $id:\phi$, where $id$ is a rule identifier and $\phi$ a ground atom, or $\bot$.
The inference expression is positive iff $\phi$ is positive and negative iff $\phi$ is negative, and inconsistent otherwise.
\end{mydef}

An inference expression includes the program $\Delta$ from which an identified atom can be deduced by an inference system.
By contrast with a Horn-clause logic system, the program $\Delta$ is not fixed and can vary because of the assumptions in the implications.
In \cite{sae13c-ictai13}, the definition of such an inference system can be found by using the adapted notion of inference expression in Definition \ref{def:inference-expression} above.
%
%
%
%
%
%
%
Solving an embedded implication amounts to add all the rules in its premise to the given program $\Delta$:

\begin{center}
$\infer[]{\Delta \vdash R_1 \land \ldots \land R_n \Rightarrow \phi}{\Delta \cup \{R_1, \ldots, R_n\} \vdash \phi}$
\end{center}

\noindent where a rule such this in the inference system is read as: If the formulas above the line can be inferred, then those below the line can also be inferred.
Also, recall that an inference expression can include either a regular atom or a restricting atom.
In the first case, we refer to such an axiom as a {\em positive axiom}, and, in the second case, as a {\em restricting axiom}.
So, $\Delta \vdash id:A$ is a positive axiom if $A$ is a regular atom, and a restricting axiom if $A$ is a restricting atom.
%
%
Like all Gentzen-style inference systems, 
$d_s : \mathcal{A} \rightarrow \mathcal{A}$ enjoys monotonicity, idempotence and inflationaryness \cite{sae13c-ictai13}, where $\mathcal{A}$ denotes the set of inference expressions for programs.


The positive information of a set of axioms is defined as follows:

\begin{mydef}{Positive Information of a Set of Inference Expressions}
    \label{def:positive}
    The positive information of the set of inference expressions $\mathcal{A}$ is the set of each axiom including a regular atom in $\mathcal{A}$ excepting those with a counterpart restricting axiom (i.e., including a restricting atom) in $\mathcal{A}$ for the same context $\Delta$.
\end{mydef}

\noindent For example, $\Delta \vdash id_1:-take(tony,eng)$ is the counterpart restricting axiom of $\Delta \vdash id_2:take(tony,eng)$.

The negative information is deduced by applying the closed world assumption (CWA) \cite{Ullman88} to inference expressions.
However, due to the restricting atoms in inference expressions, this negative information is extended with such expressions, as it is defined next:

\begin{mydef}{Closed World Assumption of a Set of Inference Expressions}
\label{def:cwa}
The closed world assumption of the set of inference expressions $\mathcal{A}$ (written as $cwa(\mathcal{A})$) is the union of the positive information in $\mathcal{A}$ and the negative inference expression for $\Delta \vdash \phi$ such that either $\Delta \vdash \phi \notin \mathcal{A}$ or $\Delta \vdash \phi \in \mathcal{A}$ where $\phi$ is a restricting axiom.
\end{mydef}

This captures the negative information which can be deduced from a couple of sources: First, the intensional notion of negative information due to the classical closed world assumption and, second, the extensional (explicit) information due to the restricting part of the program (i.e., the set of restricting axioms).
This last one is called {\em restricting meaning} from now on.

Definition 13 in \cite{sae13c-ictai13} describes the unified stratified semantics as the bottom-up construction of the semantics, stratum by stratum, in which the inductive step $\mathcal{A}^{s+1} = cwa(d_{s+1}(\mathcal{A}^{s}))$ for $s \geq 0$ builds the semantics of the database in a finite number of steps (the number of strata is finite and no function symbols are allowed).
%
%
%
The meaning of a goal $\phi$ w.r.t. a set of axioms $\mathcal{A}$ 
is defined as 
$solve(\phi,\mathcal{A}) = \{\Delta \vdash id:\psi \in \mathcal{A} ~ \mbox{such that} ~ \phi\theta = \psi\}$
where $\phi$ is a goal, $solve$ returns a bag, and $\theta$ is a substitution.
%

\section{Implementation}
\label{sect:implementation}
Last section has introduced an operational semantics which builds the semantics of the whole database in a purely bottom-up fashion.
Here, we recall some implementation details from \cite{sae13c-ictai13} and adapt it to support negative assumptions and restricted predicates.
So, we consider a top-down-driven, bottom-up fixpoint computation with tabling as implemented in the deductive system DES \cite{DBLP:conf/aplas/Saenz-PerezCG11}, which follows the ideas found in \cite{Dietrich87,TS86}.
This system is implemented in Prolog and incorporated hypothetical Datalog in version 3.2 (February, 2013) for the first time\footnote{Release notes in \url{des.sourceforge.net} lists all its history.}.
Version 3.6 (March, 2014) enhanced this by allowing negative assumptions as well as the dynamic construction of the PDG and stratification.
Next we describe implementing tabling, negative assumptions in premises, and some optimizations.

\subsection{Tabling}
\label{subsect:tabling}

Though there have been some works regarding implementations \cite{DBLP:conf/lpar/VieilleBKLM92}, as far as we know there has not been an implementation of hypothetical Datalog based on tabling and
allowing both embedded implications and stratified negation.
Tabling faces some well-known problems of logic programming implementations: 
Unsoundness, repeated computations, and termination, providing some overcomes, and it has been useful in particular for implementing efficient systems.
It has been applied to different fields (logic programming systems \cite{Guzman:2008:ICC:1785754.1785768,DBLP:journals/tplp/SwiftW12}) and in particular to deductive databases (e.g., \cite{Shen02slt-resolutionfor,sae13a}).

Systems implementing tabling memorize the deduced instances (answers) to goals (calls) in an answer table and call table, respectively, in order to reuse already available deductions.
A call table $ct$ stores the goal calls made along resolution
, and an answer table $at$ stores (ground) answers.

The inference rule for hypothetical goals defined in Section \ref{sect:formal} amounts to try to prove a goal in the context of the current database augmented with the premise of the implication.
As a literal can be of the form $R_1 \land \ldots \land R_n \Rightarrow \phi$, where $R_i$ are rules and $\phi$ a goal, the database $\Delta$ for which this hypothetical literal is to be proven must be augmented with $\{R_1, \ldots, R_n\}$.
Deductions delivered in proving $\phi$ are only valid in the context of the augmented database, i.e., in the tabling tree constructed for $\phi$.
So, such deductions must be tagged in order to be only used in its context (contexts will be denoted by $\chi$).

Filling the answer and call tables is due to the so-called {\em memo function} which proceeds by tabled SLDNF resolution as detailed in \cite{sae13c-ictai13}.
The memo function $memo(\phi,\Delta,\chi,ct,at)$ is applied, respectively, to a goal $\phi$, a program $\Delta$, a context identifier $\chi$, and input call and answer tables $ct$ and $at$, and returns the (possibly) augmented call and answer tables $ct'$ and $at'$.
An entry in the answer table has the form $id^\chi:A$, where $id$ is the program rule identifier in the context $\chi$, and $A$ is either a positive or negative or restricting atom.
The positive information of an answer table is the set of all its entries $id^\chi:A$ such that $A$ is a regular atom.
The answer table is augmented with the head of a program clause with the corresponding substitutions derived from proving each of its literals in the body clause.
A literal can be proven if an atom (either regular or restricting) is found in the closed world assumption of the input answer table for the current context.
%
%
Such closed world assumption of an answer table is defined analogously to the closed world assumption of a set of inference expressions:

\begin{mydef}{}\hspace*{-2mm}{\em (Closed World Assumption of an Answer Table):~~}
\label{def:hyp_cwa_at}
The closed world assumption of an answer table $at$ (written as $cwa(at)$) in the context of a program is the positive information of $at$, and any $\epsilon^\chi:\neg A$ such that, either $id^\chi:-A \in at$ or $id^\chi:A \notin at$ for any rule identifier $id$ and context $\chi$, where $\epsilon$ is a fixed, arbitrary identifier which does not occur in the program.
\end{mydef}

Filling the answer and call tables is done by strata by ensuring that the meaning of negated atoms which are required to prove other goals are already stored in the answer table.
So, following the stratification for the program for a given goal $\phi$, a goal dependency graph is computed, which is the subgraph of the PDG such that contains all the reachable nodes from $\phi$.
Then, for each node $p_i$ in the subgraph such that there is a negative arc coming out from $p_i$, an open goal $\phi_i$ is built with the same arity as $p_i$.
Goals $\phi_i$ are ordered by $str(\Delta,\phi_i)$, so that lower-strata goals will be computed before upper-strata goals.
The goal dependency graph is specified in \cite{sae13c-ictai13} as the function $gdg(\Delta,\phi)$ which is applied to a program $\Delta$ and goal $\phi$, returning the pair of nodes and arcs $<N,A>$.

We refer here to \cite{sae13c-ictai13} for the definitions of the stratified meaning of a program restricted to a goal (Definition 19), the fixpoint of the database built with $\bigsqcup_{n\geq 0}$, and the meaning of a tabled goal (Definition 20).

\subsection{An Example}

Following an analogous example to the one in Section III.{\em B} in \cite{sae13c-ictai13}:

\begin{tabular}{l}
$route(X,Y) \leftarrow connected(X,Y) \lor connected(Y,X)$\\
$route(X,Y) \leftarrow route(X,Z) \land route(Z,Y)$\\
$no\_route(X,Y) \leftarrow station(X) \land station(Y) \land \neg route(X,Y)$\\
\end{tabular}

\noindent for which its PDG is $<\{station,$ $connected,$ $route,$ $no\_route\},$ $\{route \leftarrow connected,$ $route \leftarrow route,$ $no\_route \leftarrow station,$ $no\_route \stackrel{\neg}{\leftarrow} route\}>$, and a stratification is $\{(station,1)$, $(connected,1)$, $(route,1)$, $(no\_route,2)\}$.
Then, let us consider in addition to this database the predicate $closed/1$ that lists stations that must be closed sometime due to workmanship.
The following rule allows to know what are the possible connections under such an assumption: 

\noindent
\begin{tabular}{rcll}
$restricted\_route(X,Y)$ & $\leftarrow$
& $(-connected(A,B) \leftarrow connected(A,B),closed(A)) \land$ \\
& & $(-connected(A,B) \leftarrow connected(A,B),closed(B))$ & $\Rightarrow route(X,Y)$
\end{tabular}

This new rule adds to the PDG the edges 
$\{connected/2 \leftarrow closed/1,$ 
$connected/2 \leftarrow connected/2,$ 
$route/2 \stackrel{\neg}{\leftarrow} connected/2,$ $restricted\_route/2 \leftarrow route/2\}$ and the stratification becomes: 
$\{(closed/1,1),$ 
$(connected/2,1),$ 
$(station/1,1),$ 
$(route/2,2),$ 
$(restricted\_route/2,2)\}$,
where both 
$restricted\_route/2$ and $route/2$ are located at the second stratum due to the negative assumption on $connected/2$ that imposes the negative arc $route/2 \stackrel{\neg}{\leftarrow} connected/2$.

\subsection{Implementing Tabling}

DES implements implications in Prolog as described in \cite{sae13c-ictai13}.
Recalling, each time an implication is to be solved, a new context is created by augmenting the current database with the rules and facts in the premise.
If the same program point is reached for solving the implication due to the fixpoint computation (corresponding to a new iteration), then the database is not changed because the program rules for the premise are already loaded and tagged for that context.
Entries in the call and answer tables are accordingly tagged so that the outcome for a given context can be identified as well.
Solving a goal $g$ in a stratum greater than 1 proceeds by stratified computation as described in 
\cite{sae13c-ictai13},
i.e., solving stratum by stratum the meaning of the involved predicates on which $g$ negatively depends, and solving $g$ with the results for other predicates already stored in the answer table.
Next, first the implementation of solving restricted predicates is depicted and, then, a couple of optimizations are proposed.

\subsubsection{Solving Restricted Predicates}

Solving a call to a restricted predicate $p$ is also done by stratum because its actual (restricted) meaning must be computed before any predicate that depends on $p$.
The rationale behind this solving is to compute both the positive part and the restricted part of $p$ by considering, respectively, its defining rules with regular and restricting atoms in the head.

As any predicate with an outgoing negative dependency, the restricted predicate $p$ is located at a higher stratum than each $q_i$ such that $q_i \stackrel{\neg}{\leftarrow} p$ is in the PDG.
This implies that $p$ is to be solved (before each $q_i$) in its stratum for an open call \mytt{p(X$_1$,...,X$_n$)}, where $n$ is the arity of $p$ and \mytt{X$_i$} are fresh variables.

The next code excerpt illustrates the solving of a single call (either regular, restricted or negative) for a given stratum:

{\myfontcodesize
\begin{verbatim}
solve_datalog_stratum(not Q,Stratum,CId,Undefined) :-
  solve_datalog_stratum(Q,Stratum,CId,_Undefined), !,
  solve_positive_datalog_stratum(not Q,Stratum,CId,Undefined).
solve_datalog_stratum(Q,Stratum,CId,Undefined) :-
  solve_pos_res_datalog_stratum(Q,Stratum,CId,Undefined).

solve_pos_res_datalog_stratum(Q,Stratum,CId,Undefined) :-
  solve_positive_datalog_stratum(Q,Stratum,CId,Undefined),
  functor(Q,N,A),
  (restricted_predicate(N/A)
   ->  solve_positive_datalog_stratum(-Q,Stratum,CId,_Undefined2),
       remove_restricted_tuples(Q,CId) 
   ;   true).  
\end{verbatim}
}

Here, the predicate \mytt{solve\_datalog\_stratum} is responsible of solving a given call (first argument) in a stratum (second argument).
Its third argument \mytt{CId} corresponds to the context identifier $\chi$ as introduced already.
The last argument \mytt{Undefined} stands for undefined results, which are got for non-stratifiable databases\footnote{Such behaviour is allowed for teaching purposes in order to highlight the problems in trying to compute non-stratifiable databases.}.

For a non negated call (second clause) a possibly restricted call is solved with  \mytt{solve\_pos\_res\_data-}\linebreak \mytt{log\_stratum}.
This predicate first solves the non-restricted (i.e., positive) meaning of the call \mytt{Q} and, if it refers to a restricted predicate, then its extensional negative meaning (corresponding to the restricting call \mytt{-Q}) is computed. 
After computing \mytt{-Q}, both the positive and extensional negative meanings are already stored in the answer call, and the actual meaning is changed in the answer call by removing all entries with a counterpart restricting atom.
For example, if \mytt{\{p(1),p(2),-p(2)\}} are in the answer table after solving the calls \mytt{Q} and \mytt{-Q} for a given context $\chi$, then the resulting meaning for \mytt{p} is just \mytt{\{p(1),-p(2)\}}, where \mytt{p(2)} has been removed by \mytt{remove\_restricted\_tuples}.
So, given this answer table for a context $\chi$, the call \mytt{-p(1)} does not succeed because  \mytt{-p(1)} is not in the restricting meaning of \mytt{p} for $\chi$. 
As well, the call \mytt{-p(2)} succeeds and  the call \mytt{-p(3)} does not succeed for analogous reasons.

For a negative call \mytt{not Q} (first clause), \mytt{Q} is firstly solved as before.
This fills the answer table with the (possibly restricted) meaning of \mytt{Q}, so its negation can be solved with the call \mytt{solve\_positive\_data-}\linebreak \mytt{log\_stratum(not Q,\ldots)}
The negative meaning of \mytt{Q} is composed of its extensional negative meaning (restricted part of the predicate) and its intensional negative meaning (which follows SLDNF).
So, continuing with the last example, the calls \mytt{not -p(1)}, \mytt{not -p(2)}, and \mytt{not -p(3)} respectively succeeds, does not succeed, and succeeds.
Note that the call \mytt{not p(3)} succeeds (as \mytt{not -p(3)} does) because {\tt p(3)} cannot be proven by SLDNF.
This is equivalent to say that, with the available information, neither \mytt{p(3)} nor \mytt{-p(3)} can be proved.
Finally, it is not possible to solve \mytt{not -p(X)} simply because the query is unsafe (c.f., Section II.{\em B} in \cite{sae13c-ictai13}).

A context can be thought of as the current database along query solving which has been modified with respect to the original database due to positive and negative assumptions (i.e., by respectively adding regular and restricting rules).
A na\"ive implementation of contexts would be to represent each rule of the current database in the parameter $\chi$.
Instead, we resort to tag each program rule with a context identifier, which is identified as a list of integers.
Each integer in this list corresponds to the rule identifier in which an assumption is made.
The initial context is the empty list, and only entries in the answer table referring to this context are kept, though along the computation, entries for other contexts are kept.
Let us consider the following simple program (where each rule is identified by an integer and a context between parentheses):

{\myfontcodesize
	\begin{verbatim}
(0,[]): p :- q => r.
(1,[]): r :- q.                         
	\end{verbatim}
}

After solving the query \mytt{p} (which succeeds), the answer table includes the (simplified) tuple \mytt{(p,[])}, indicating that \mytt{p} is true in the initial context.
For solving this query, an assumption is made, which amounts to locally adding the fact \mytt{q} (as a new rule identified by the integer \mytt{2}) to the initial context.
This addition is implemented as the assertion of this new rule for the context \mytt{[0]} becoming:
{\myfontcodesize
	\begin{verbatim}
(2,[0]): q.
	\end{verbatim}
}

The new rule is asserted only once along fixpoint iterations, and it is removed at the end of query solving.
When assumptions are nested, as in \mytt{p :- q => r => s}, the rule is transformed by removing nested assumptions:
{\myfontcodesize
	\begin{verbatim}
(0,[]): '$p0' :- r => s.
(1,[]): p :- q => '$p0'.
	\end{verbatim}
}
Here, the new predicate \mytt{\$p0} is automatically created during preprocessing so that each context can be identified by a single rule.

\subsubsection{Dynamic PDG and Stratification}

Section \ref{sect:pdg} introduced the construction of the predicate dependency graph by considering all the rules defined in the program, including those in the antecedent of embedded implications.
However, a more refined approach can be considered by building a PDG and stratification for each context.
Indeed, rules that do not form part of a given context may introduce negative dependencies which imply to solve the {\em complete} meaning of a given predicate, instead of considering only its actual context.
For instance, let us consider the rules $p(X) \leftarrow t(X)$ and $q(X) \leftarrow (p(Y) \leftarrow t(Y) \land \neg r(Y)) \Rightarrow s(X)$.
The arcs in the corresponding PDG are $\{ p/1 \stackrel{\neg}{\leftarrow} r/1, p/1 \leftarrow t/1,  q/1 \leftarrow s/1\}$, and a possible stratification is $\{(q/1,1),$ $(r/1,1),$ $(s/1,1),$ $(t/1,1),$ $(p/1,2) \}$.
So, consider the goal $p(1)$, whose solving  proceeds by stratum: First, the open goal $r(X)$ is solved in the first stratum (providing the complete meaning of $r/1$ in the answer table), then the goal $p(1)$ is solved in the second stratum (no other predicates are considered since the computation is restricted to the goal dependency graph).
But consider that $r$ can contain millions of tuples, and all of them are computed when they are not really needed.
If only one tuple of $t$ matched the call for $p$ then only one tuple would be needed.
The negative dependency that forces $p$ to be in stratum 2 comes from a premise that is not involved in the current solving.
So, we build a specific (dynamic) predicate dependency graph and stratification for each context, which are correspondingly tagged and therefore avoids such wasteful computations.
Thus, 
the arcs in the PDG and stratification for the goal $p(1)$ are, respectively, $\{ p/1 \leftarrow t/1,  q/1 \leftarrow s/1\}$, and all nodes remain in a single stratum.
Solving $p(1)$ fills only one tuple of the answer table for $t/1$ and no one for $r/1$.

In the concrete implementation, this program is written as follows (where identifiers have been included as before):
{\myfontcodesize
	\begin{verbatim}
(0,[]) p(X) :- t(X).
(1,[]) q(X) :- (p(Y):-t(Y), not r(Y)) => s(X).
	\end{verbatim}
}
The PDG and strata for the top-level context \mytt{[]} can be inspected with:
{\myfontcodesize
	\begin{verbatim}
DES> /pdg
Nodes: [p/1,q/1,s/1,t/1]
Arcs : [p/1+t/1,q/1+s/1]
DES> /strata
[(p/1,1),(q/1,1),(s/1,1),(t/1,1)]
	\end{verbatim}
}
\noindent where \mytt{P+Q} (\mytt{P-Q}) denotes that the predicate \mytt{P} positively (negatively, resp.) depends on the predicate \mytt{Q}.

When solving the query \mytt{q(X)}, the following PDG and strata are computed for the new context \mytt{[1]} due to the assumption in rule \mytt{(1,[])}, which can be displayed by enabling verbose output (with \mytt{/verbose on}).
{\myfontcodesize
	\begin{verbatim}
DES> q(X)
...
Info: Building hypothetical computation context [1] for:
p(Y) :- t(Y), not r(Y).
Info: PDG:
Nodes: [p/1,q/1,r/1,s/1,t/1]
Arcs : [p/1-r/1,p/1+t/1,q/1+s/1]
Info: Strata:
[(q/1,1),(r/1,1),(s/1,1),(t/1,1),(p/1,2)]
...
	\end{verbatim}
}
The PDG and strata are incrementally built for each modification (rule addition or deletion) in the database.
So, when an assumption is made, they are updated according to the assumption (recall that an assumption always adds a rule, either regular or restricting).
Upon entering into a new context, the old PDG and strata are saved and eventually restored when the computation for the new context is finished.


\subsubsection{Reusing Answers from Previous Contexts}
\label{sect:optimization}
Solving an embedded implication as presented requires to recompute from scratch the given goal for all the involved strata.
While this is a conservative approach, former computations in previous contexts can be reused to avoid some recomputations, i.e., reusing entries in the answer table. 
For the database restricted to the goal 
consisting only of a single stratum, this reusing is safe as only additions to the answer table are possible.
So, retrievals from the answer table can be done from the first context up to the current one.
However, when negation is involved in this restricted database, some already deduced information in a former context might be not true anymore.
Consider, for instance, the program consisting of the identified rules \mytt{(1,[]):} $p \leftarrow \neg q$ and \mytt{(2,[]):} $r \leftarrow q \Rightarrow p$.
The goal $p$ succeeds in the initial context \mytt{[]}, but fails in the context \mytt{[2]} when solving the conclusion $p$.
A straightforward implementation for facing this issue is simply to avoid the reusing for strata greater than 1, which can be done by adding a new parameter to the predicates stating the current stratum.
Another, more refined implementation is by identifying those predicates which do not depend on assumed information, either directly or indirectly, and avoiding the reusing of their deduced information, committing only to the current context.

\section{Conclusions and Future Work}
\label{sect:conclusions}
This work has presented a novel add-on to deductive databases: hypothetical rules with negative assumptions in the premise of embedded implications, extending both \cite{bonner90adding} and \cite{sae13c-ictai13}.
The work \cite{bonner90adding} has been extended with restricting rules in the premise (not only facts), retaining also the extensions in \cite{sae13c-ictai13} (duplicates and strong constraints).
Also, \cite{sae13c-ictai13} has been extended by providing the novel concept of restricted predicates as a means to prune the meaning of predicates due to negative assumptions.
In addition, a dynamic construction of the PDG has been proposed as well as another optimization for pruning computations.
We have described an implementation for our proposal as part of  the publicly available system DES.
Since SQL queries in DES are translated into Datalog rules,  this technique also supports negative assumptions in SQL queries (Version 3.10, January 2015).
As future work, first we envision to implement the optimization for pruning computations.
Performance data can be taken to highlight the gains of the proposed optimizations.
Second, it should not be hard to devise the non-encapsulated vision of premises, by setting the scope of variables in the premise to the whole rule or goal in which it occurs.
Finally, we are currently widening the semantics and implementation for allowing guessing in premises, i.e., to infer the hypothetical data in the antecedent to prove a given consequent.


\end{document}